\documentstyle[epsf,epsfig]{disk}
\begin{document}

\title{%
HST Observations of X-Ray Transients}

\author{Carole A. HASWELL\\
{\it Astronomy Centre, University of Sussex,
Falmer, Brighton\\ BN1 9QJ, United Kingdom, 
chaswell@star.cpes.susx.ac.uk}\\
Robert I. HYNES\\
{\it Astronomy Centre, University of Sussex,
Falmer, Brighton\\ BN1 9QJ, United Kingdom, 
rih@star.cpes.susx.ac.uk}\\
}

\maketitle

\section*{Abstract}

Results from an HST/multiwavelength
campaign on the 1996 outburst of GRO J1655-40
are reported. 
We find evidence for an approximately isothermal
outer disk in GRO J1655-40 during the decline from
outburst, possibly indicating the outer disk being
maintained in the hot state.
HST observations of 
GRO J0422+32 made 2 years after outburst
reveal a sharply peaked optical/near UV spectrum,
providing strong evidence for non-thermal emission.

\section{Introduction}
This paper summarises results
from our HST/multiwavelength campaign on the 1996 
outburst of GRO J1655-40, and near quiescent
1994 observations of GRO J0422+32.
 Our primary goal is to probe and understand the mechanisms
causing the dramatic luminosity evolution in
these systems.

\section{GRO J1655-40}
\begin{figure}
\label{v1}
\vspace{-0.9truein}
\epsfysize=2.5in\epsfbox[20 170 530 580]{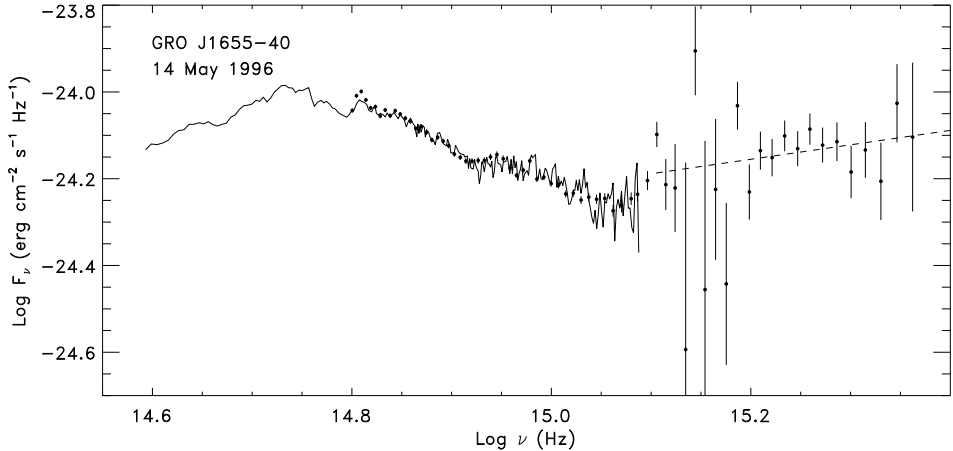}
\vspace{0.9truein}
\caption{The dereddened spectrum of GRO~J1655-40. The spectral slope
changes dramatically at 
${\rm log ~ \nu \sim 15.05}$ (${\rm \lambda \sim 2600 \AA}$).
The dashed line shows a ${\rm f_\nu \propto \nu ^{1/3}}$
fit to the UV data. Adapted from Hynes et al. 1998a.}
\end{figure}
The superluminal jet source GRO~J1655-40 
has undergone repeated outbursts
since its 1994 discovery.
We mounted a coordinated
HST and RXTE campaign
between
1996 May~14
and July~22.
Fuller descriptions of these observations and their interpretation
are given
in Hynes et al. (1998a,b).

\subsection{The 1996 Broad Band Spectral Evolution of GRO J1655-40}
Modelling of the ${\rm 2175\AA}$ feature
yields E(B-V)=$1.2\pm0.1$
(Hynes et al. 1998a).
Figure~1 is the 1996 May~14 dereddened UV-optical spectrum.
Though the UV portion of the spectrum is consistent
with the $\nu ^{1/3}$ power-law predicted by
the steady-state blackbody disk model, the optical
(${\rm \lambda > 2600 \AA}$) spectrum rises to longer wavelengths
in contrast to the predictions of the model.
Ignoring the ${\rm \lambda >  2600 \AA}$ data, a $\nu ^{1/3}$ model
can be fit to the UV data, implying a
mass transfer rate of
${\rm 1\times 10^{-7}}$\,M${\rm _{\odot}}$\,yr${\rm ^{-1}\leq
 \dot{M}\leq
 7\times 10^{-6}}$\,M${\rm _{\odot}}$\,yr${\rm ^{-1}}$,
if the compact object mass is ${\rm 7~M_{\odot}}$ as
the quiescent observations imply (Orosz and
Bailyn 1997, hereafter OB97).
The dominant source of uncertainty in this ${\rm \dot{M}}$ arises from
the extinction correction.
Assuming an accretion efficiency
of ${\rm 10 \%}$, the Eddington rate is
 ${\rm \dot{M}_{\rm Edd}=1.6\times 10^{-7}}$\,M${\rm_{\odot}}$\,yr${\rm ^{-1}}$,
so near the peak of the outburst this interpretation of the UV spectrum implies
${\rm \dot{M} \approx \dot{M}_{\rm Edd}}$.

We need to invoke something other than a pure steady-state optically thick
accretion disk to explain the optical light. The shape
of the spectrum is qualitatively suggestive of an irradiated disk; irradiation
can alter the temperature profile of the outer disk producing a rise
in flux
towards longer wavelengths as illustrated in Fig.~2a.
The multiwavelength light curves for the outburst (Hynes et al. 1998a)
do not 
support a {\it simple} irradiation model: the optical and UV flux declines
while the X-ray flux rises.
Nontheless correlated X-ray and optical/UV variability
was detected (see section 2.3 and Fig. 3a)
indicating at least some of the optical/UV flux is due to
reprocessing in the disk.
%
%
%

While the optical fluxes fell by about a factor of three between
our first and last visit, the colour temperature remained almost
constant, dropping
from 9800~K to 8700~K; the effective
emitting area dropped from ${\rm 5.0 \times 10^{23}~c
m^2}$
to ${\rm 2.2 \times 10^{23}~cm^2}$ (Hynes et al. 1998a.)  
The well constrained system parameters for GRO~J1655-40
(OB97, Hjellming and Rupen 1995)
imply the total available emitting area 
is ${\rm \sim 5 \times 10^{23}~cm^2}$, so it is possible to explain
the optical emission at the peak of the outburst as thermal emission,
but only if both the secondary star and the majority of the disk area have
essentially the same temperature.
In this scenario, 
the $\nu ^{1/3}$
UV component can, however, still be attributed to 
the hot inner regions of the disk.
 
 
While noting that the spectral shape is strongly
dependent on the adopted reddening correction, our optical spectra appear
more strongly peaked than a single temperature blackbody,
so we considered non-thermal emission mechanisms (Hynes et al.
1998a, especially Fig. 7 and 11).
Self-absorbed
synchrotron emission from a compact cloud of relativistic electrons
produced good fits to the optical component.
Attributing substantial optical
flux to this compact non-thermal source relieves the requirement
for a large isothermal emitter in the system. 
But while the synchrotron models fit better than black bodies, they have 
an extra free parameter,
and there is no external check comparable to that provided by the emitting area
required for the black body interpretation.
The intrinsic VRI band linear polarisation  ($> 3 \%$)
detected in July 1996 (Scaltriti et al. 1997) seemed consistent
with the hypothesis of optical synchrotron emission.
Subsequent phase-resolved polarisation measurements
(Gliozzi et al. 1998)
seem, however, to indicate an extended polarisation region, 
inconsistent
with the possible compact non-thermal source.
The lack of optical dips during near-total X-ray dips
(Fig. 3b) provides further evidence against
a compact central source for the optical
continuum.
Hence, on balance a thermal origin for the optical continuum
seems most likely.
 
Indeed, the DIM may {\it require} a large isothermal
outer disk for a long-period system like GRO~J1655-40.
Since the disk in such a system is large, the temperature
for a steady state disk falls below
the minimum temperature for the hot, high
viscosity, state long before the outer disk is reached.
Even for an Eddington mass transfer rate in GRO J1655-40
the temperature drops below that of the hot state at
a radius less than a quarter of
the Roche lobe radius.
This means that there is no global steady-state solution for
${\rm \dot{M} \leq \dot{M}_{Edd}}$. It is not clear what will
happen to the temperature distribution in such
a case, but it is possible that in outburst much of the outer
disk could be maintained just
in the hot state. Hence one might expect the outer disk to
appear as an approximately
constant temperature, shrinking area emitter as the decline
proceeds. This hypothesis awaits tests with self-consistent
numerical modeling of the DIM in long period systems. 
 
\begin{figure}
\label{pcyg}
\begin{center}
{\bf a)}
\hspace{-10mm}
\epsfig{width=2.1in,angle=90,file=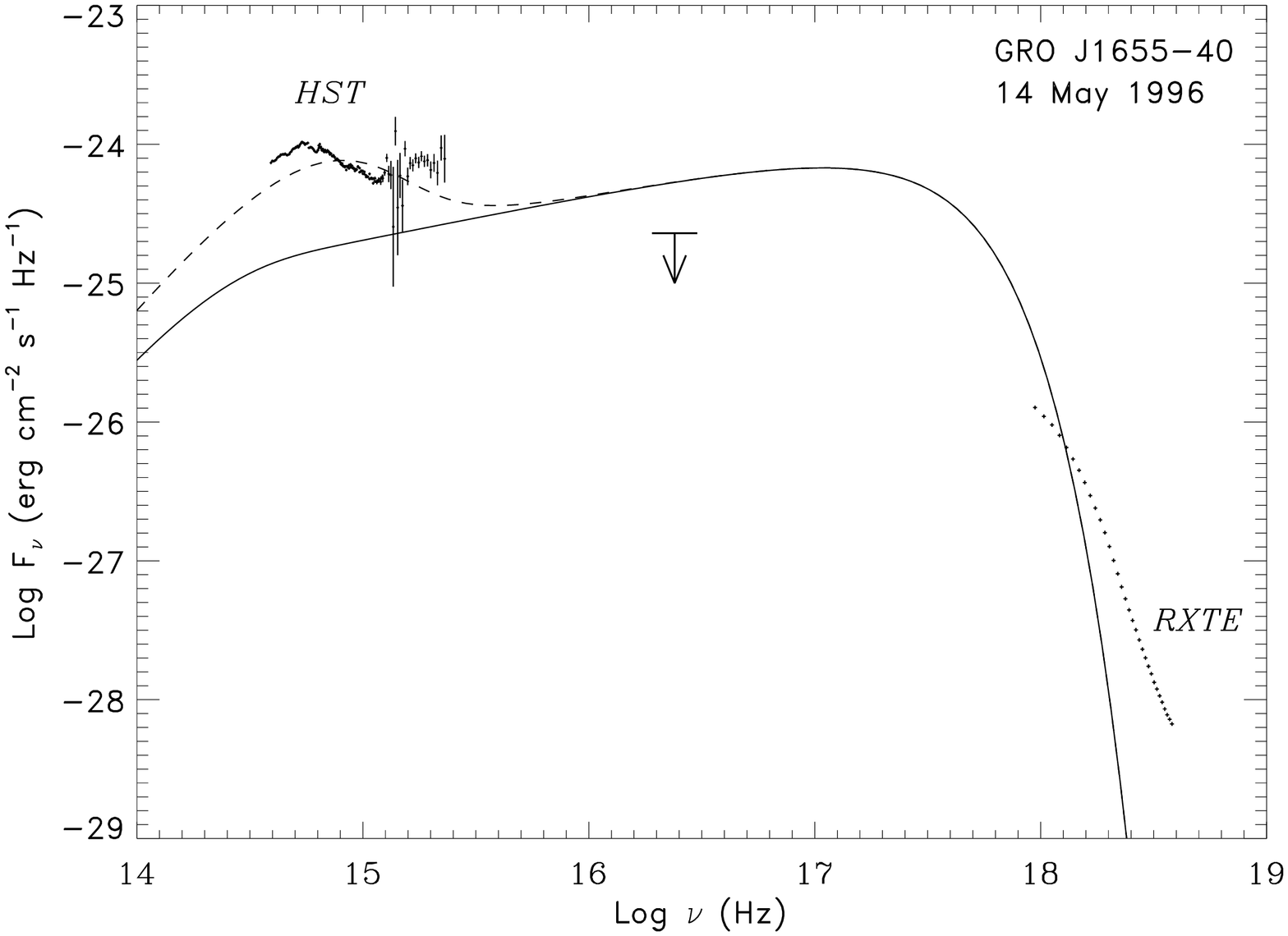}
\hspace{\fill}
{\bf b)}
\hspace{-10mm}
\epsfig{width=2.1in,angle=90,file=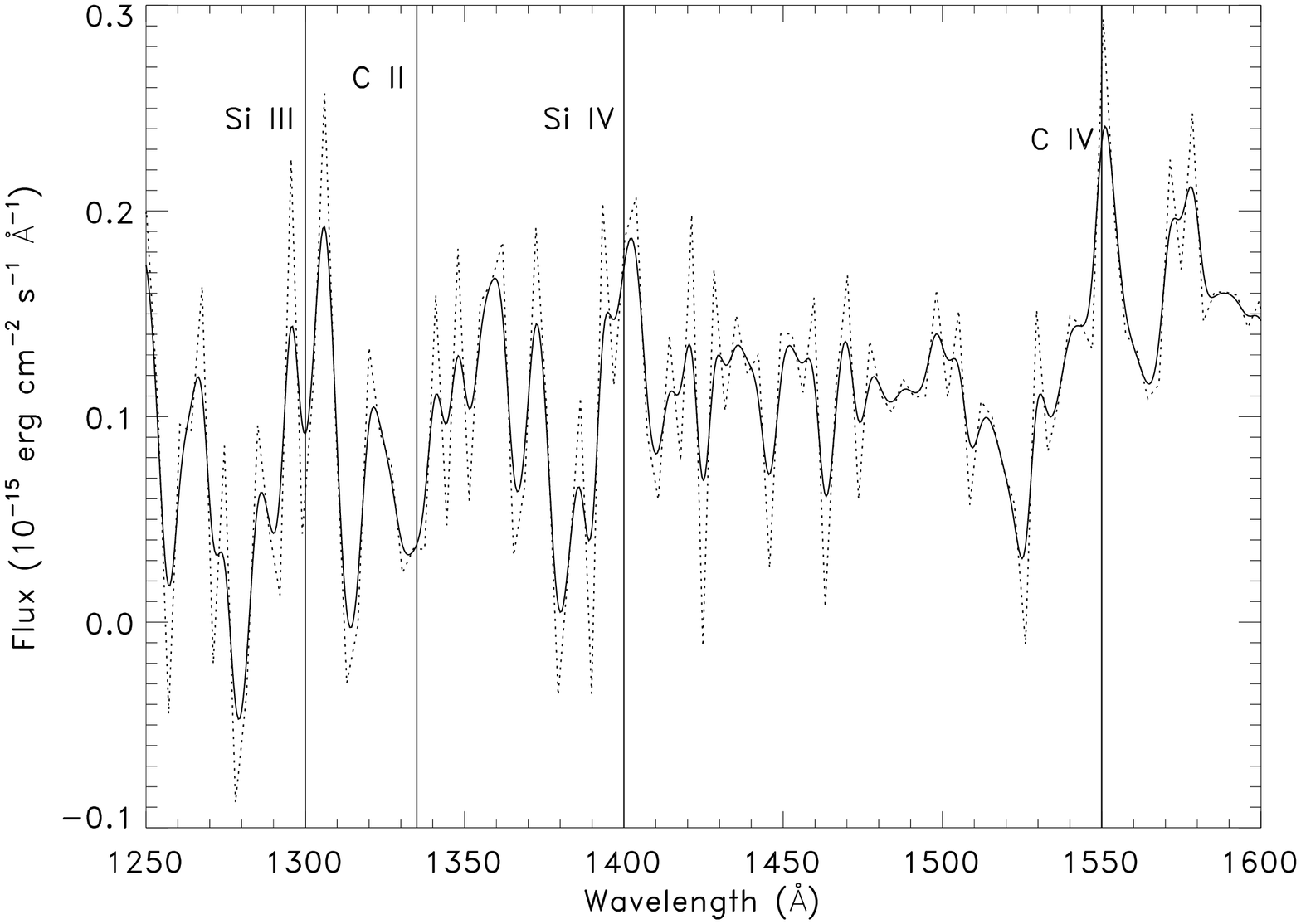}
\caption{a) Composite spectrum for the May 14 observations. The solid
line shows a steady state accretion disk spectrum; the dashed line
illustrates a simplistic model including irradiation. These model
spectra are 
{\bf not} fitted to the data, and are merely illustrative.  b) Likely
P-Cygni profiles detected in the UV resonance lines in the 1996 May 14
UV spectrum of GRO J1655-40. The dotted line shows the raw data; the
solid line is the same smoothed with a width equal to a resolution
element. From Hynes et al. 1998a.}
\end{center}
\end{figure}

\subsection{A Disk Wind?}
We found likely P-Cygni profiles  in the
UV resonance lines (Fig~2b).
The peak to trough separation is ${\rm \sim 5000~km~s^{-1}}$,
slightly larger than seen in outbursting dwarf novae.
Line profiles produced by
biconical accretion disk winds were
calculated by Shlosman and Vitello (1993) who found `classical' P~Cygni profiles
only for inclinations around 60--70$^{\circ}$, in
striking agreement with the inclination
determined for GRO~J1655-40 (OB97).
We conclude that there was likely a biconical accretion disk wind
at the peak of the UV outburst, when ${\rm \dot{M} \sim \dot{M}_{Edd}}$.
\begin{figure}
\begin{center}
{\bf a)}
\hspace{-10mm}
\epsfig{width=2.1in,angle=90,file=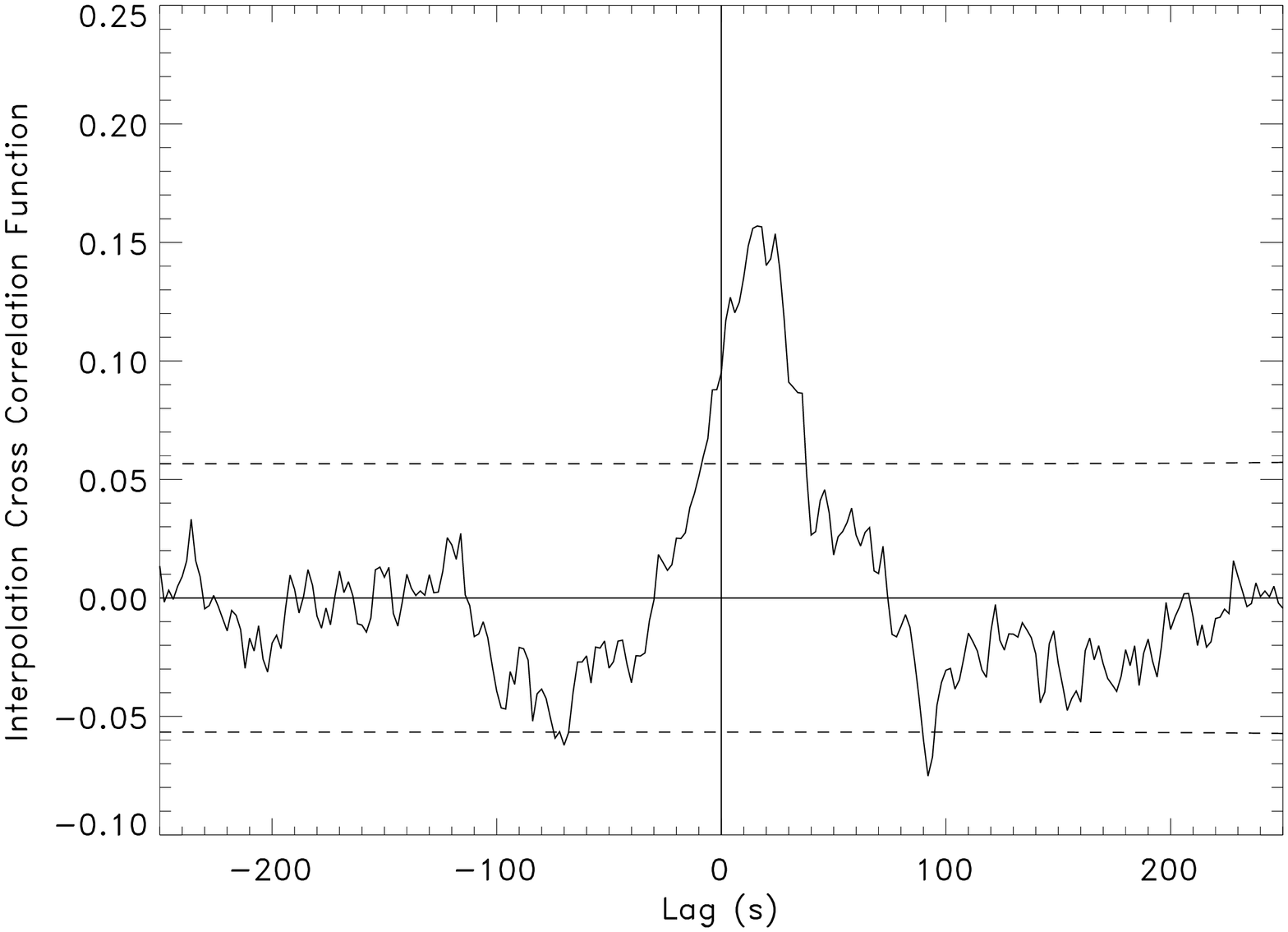}
\hspace{\fill}
{\bf b)}
\hspace{-10mm}
\epsfig{width=2.1in,angle=90,file=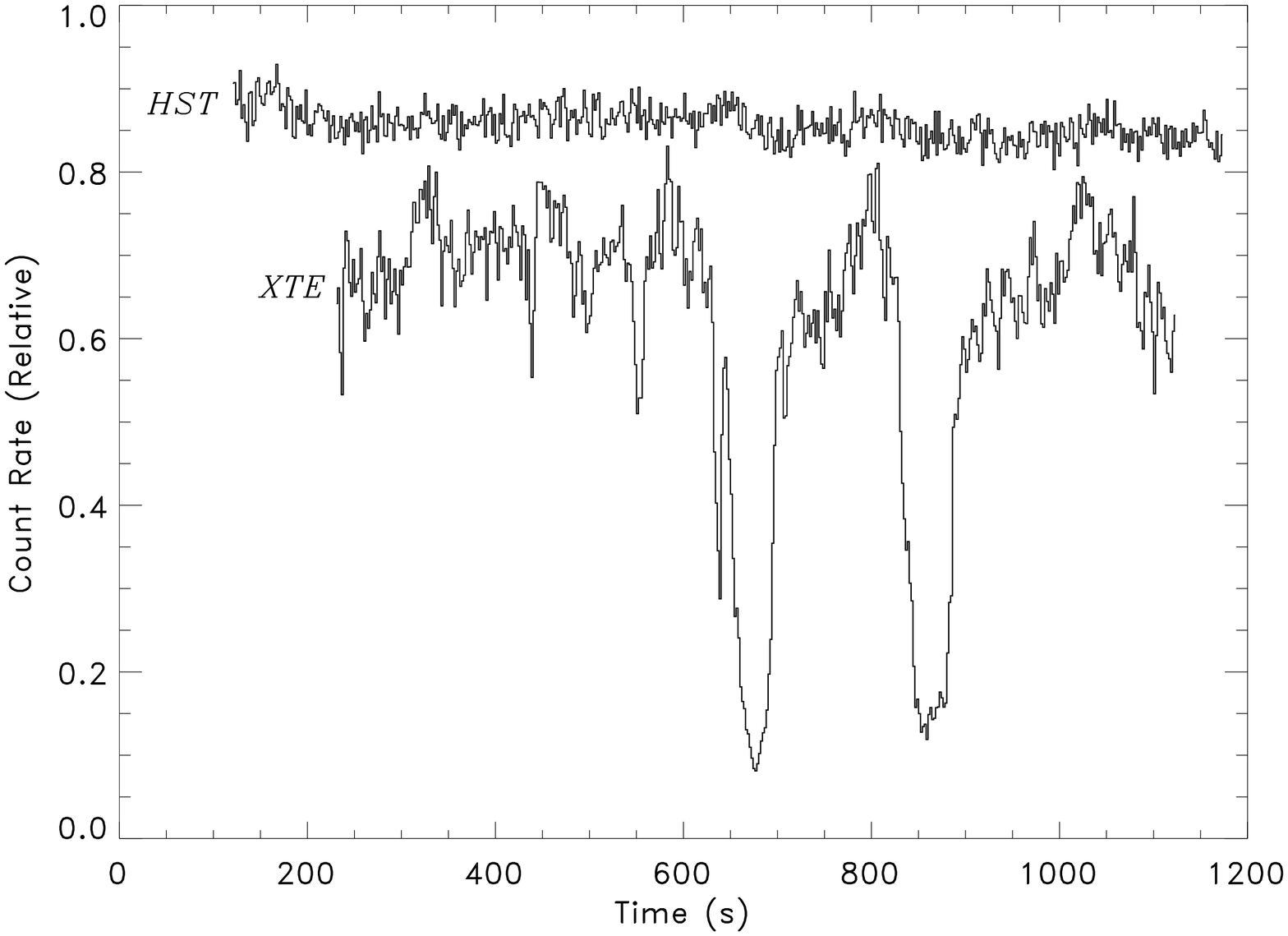}
\caption{a) The combined interpolation cross-correlation function of
our 1996 June 8 {\it HST/RXTE} observations of GRO~J1655--40.  Dashed lines
denote $3\sigma$ limits on spurious correlations.  There is a clear
correlation with a lag $\sim20$-sec.  b) {\it HST} and
{\it RXTE} light curves from 1996 June 20, showing two X-ray dips, but
no corresponding optical dips.}
\label{MultiICFFig}
\end{center}
\end{figure}
\subsection{Light Echoes}
Correlated rapid variability between the
optical/UV and
X-ray emission  occurred 
in simultaneous RXTE and HST
data from 1996 June 8 (Fig.~3a). The mean delay
of the optical/UV was
up to 25\,s (Hynes et al. 1998b).
Hence, the correlations are likely due to reprocessing
of the X-rays into optical and UV emission, 
the delay being due to the finite light travel
time between the X-ray source and the reprocessing regions.
The lag of up to 25\,s is consistent with reprocessing
in the accretion disk. At the binary phase of the observations,
lags of greater than 40\,s are expected for light echoes from the
mass donor star. 
The lack of a reprocessing signal from the mass donor
star
may imply that the X-ray absorbing material
in the accretion flow is sufficiently vertically extensive
to effectively shield it from direct X-ray illumination.
For the Roche geometry deduced by OB97, this implies ${\rm H/R}$
is at least ${\rm \sim 0.25 }$ along the line of centres.

\begin{figure}
\begin{center}
{\bf a)}
\hspace{-10mm}
\epsfig{width=2.1in,angle=90,file=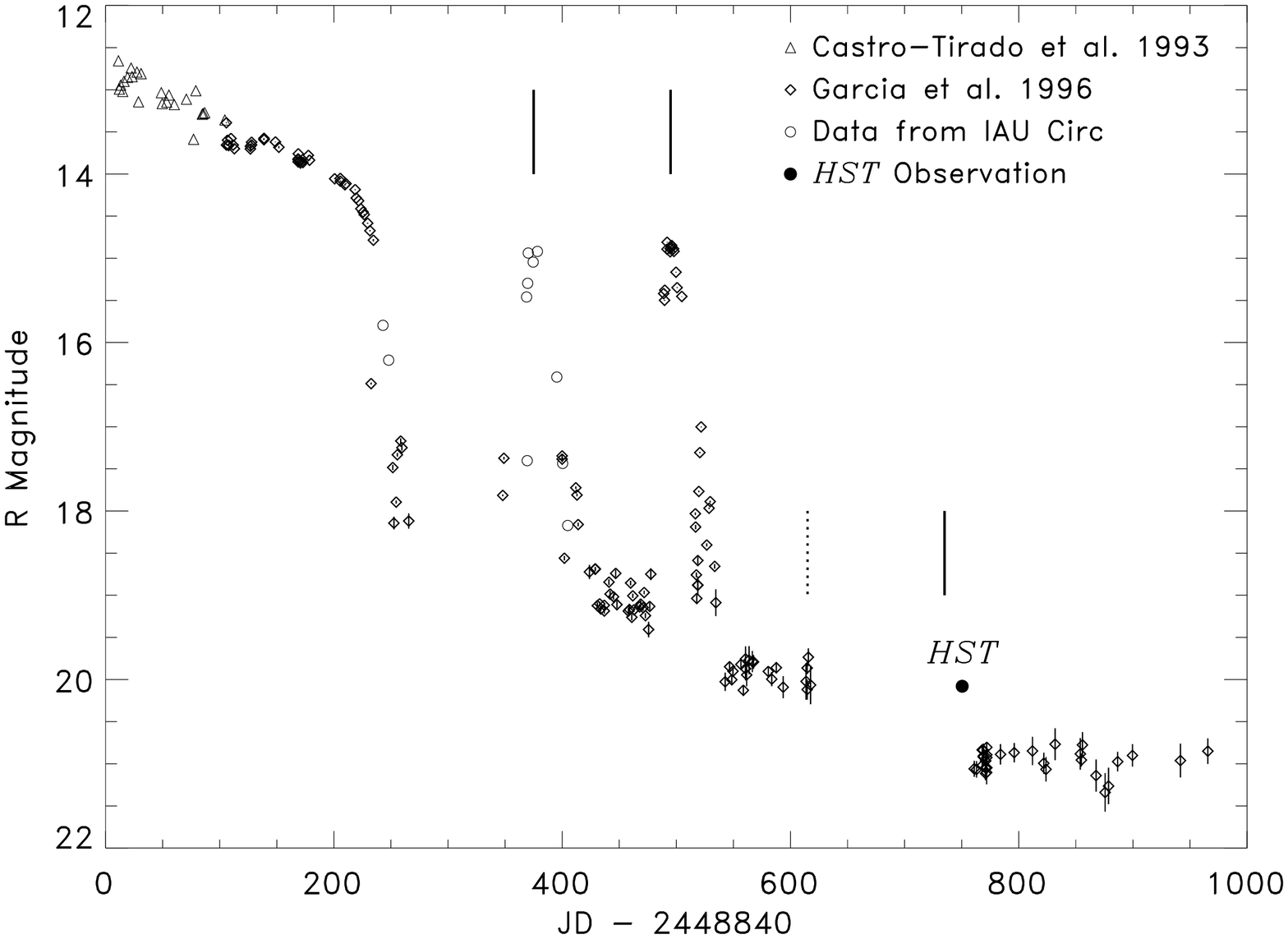}
\hspace{\fill}
{\bf b)}
\hspace{-10mm}
\epsfig{width=2.1in,angle=90,file=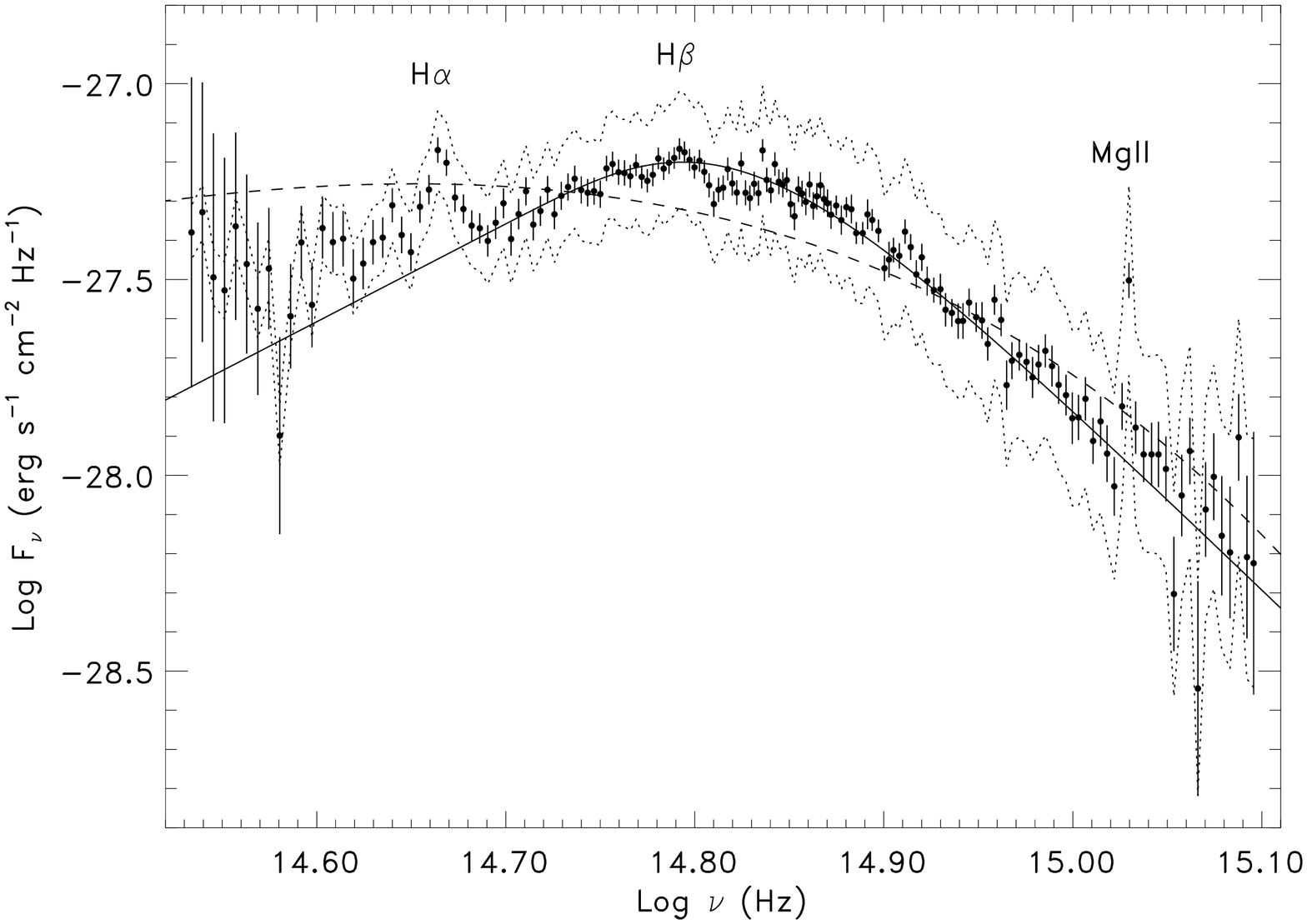}
\caption{a) The R band light curve of the outburst of GRO~J0422+32,
adapted from Fig.\ 1 of Garcia et al.\ (1996).  Vertical marks
indicate observed or extrapolated times of mini-outburst.  Our
measurement clearly lies significantly above the subsequent
photometry, and is consistent with the expected time of a
mini-outburst.  b) The near quiescent accretion spectrum from
GRO~J0422+32: dereddened data after subtraction of the companion
contribution.  The best fitting black body (dashed) and self-absorbed
synchrotron (solid) spectra are overplotted.  The data below
4000\,\AA\ ($\log\nu>14.88$) has been averaged into 25\,\AA\ bins for
clarity.  The large error bars at low frequencies are systematic and
arise from assuming a $\pm1$ pixel miscentring uncertainty.  
Dotted lines indicate the effect of varying the assumed E(B-V) value
by $\pm 0.1$.
From
Hynes and Haswell 1999.}
\label{PhotFig}
\end{center}
\end{figure}
\section{GRO J0422+32}
HST observations of this target covering the vacuum UV and the entire
optical region were obtained on 1994 August 25 and 26,
approximately two years after the first observed outburst, and seven months
after the last reported reflare (Callanan et al. 1995).
A full discussion of these data is given in Hynes and
Haswell (1999).
Figure~4a shows the R band light curve including the
point deduced from our spectrum.
The vertical marks above the light curves indicate the times of mini-outbursts
according to the suggestion (Augusteijn, Kuulkers, and Shaham 1993,
Chevalier and Ilovaisky 1995)
that they recur on a 120 day period. Our
observation clearly lies above the subsequent points,
and it appears we saw the end of a previously
unreported mini-outburst or of an extended plateau.
 
Figure 4b
shows the spectrum after dereddening
and subtracting the mass donor star flux, hence
represents the intrinsic accretion spectrum.
The spectral shape in Fig.~4b is distinctive: there is
a pronounced peak in the optical at 
${\rm \sim 4500 \AA}$;
and since the interstellar reddening
towards this system is moderate, ${\rm E(B-V) = 0.3 \pm 0.1 }$,
it is unlikely that this is an artifact of an improper
reddening correction.
The best fitting black body shown in Figure 4b
is clearly less sharply peaked than the observed
spectrum from GRO J0422+32, while a simple
self-absorbed synchrotron spectrum can successfully reproduce the
continuum shape.
We stress, however, that 
this model is very simplistic and
conclude simply that the continuum
spectrum is highly suggestive of
a self-absorbed synchrotron source, rather than
black body emission.
When compared with 
advective 
models, in which the quiescent optical 
continuum is dominated by synchrotron
emission, the agreement between
model and data is encouraging (Esin 1998).

\section{Discussion and Future Work}
 
The advective
models for quiescent black hole transients
predict that the inner disk is replaced by a hot,
optically thin,
quasi-spherical flow.
Figure~4b reveals strong evidence for non-thermal
optical
emission in GRO J0422+32 near quiescence;
this
does not necessarily confirm 
particular
detailed advective models, but 
supports the general scenario.
We plan to perform detailed comparisons
of our GRO J0422+32 spectrum with the latest models.

We also considered the possibility
of non-thermal emission
in the optical light from GRO J1655-40
during the decline from its 1996 outburst.
In this case, however, it is alternatively possible to attribute
the optical spectra to thermal emission,
as the sharply peaked spectral shape is 
dependent on the exact details of the (large) 
reddening correction. 
The non-appearance of an optical modulation
during near-total X-ray dips shown in Fig~4b
suggests that the X-ray source is far more compact
than the source of the optical continuum, 
supporting the standard thermal interpretation
of the optical continuum.

At the peak of the optical/UV light curves in 1996,
the UV spectrum from GRO J1655-40 was consistent
with an Eddington-rate mass transfer through
the UV emitting parts of the disk. At this time,
however, the X-ray luminosity was less than  ${\rm \sim 0.1~L_{Edd}}$.
This apparent discrepancy may be partly due to outflow
from the inner regions of the disk: Fig.~2b shows
likely P-Cygni profiles indicating a disk wind with
velocity ${\rm \sim 5000~km~s^{-1}}$.

In both sets of observations described here, the vacuum UV spectrum
was of low signal to noise ratio. Our single UV spectrum of
GRO J1655-40 proved informative, despite the effects of the
large reddening. We have our multiwavelength network poised
to observe newly discovered transients, and anticipate
obtaining high quality UV/optical spectra through
decline. With this program we should be able to
address many of the outstanding issues regarding
the temperature structure, morphology, irradiation
behaviour, and outflow characteristics of these systems.

{\bf Acknowledgments:}
Support
for this work
was provided by the Nuffield Foundation
and by NASA through grant numbers
GO-6017.01-94A and GO-4377.02-92A
from the Space Telescope Science Institute, which is operated by
AURA
under NASA contract NAS5-26555.
We are grateful to our collaborators
for their contributions to this work.
Fuller discussions of the work
described herein is in Hynes et al. (1998a,b) and
Hynes and Haswell (1999).

\section{References}

\vspace{1pc}

\re
1.\ Augusteijn, T., Kuulkers, E.,  and Shaham, J., 1993, A\&A, 279, L13.
\re
2.\ Callanan P. J. et al., 1995, ApJ, 441, 786
\re
3.\ Chevalier C., Ilovaisky S. A., 1995, A\&A, 297, 103
\re
4.\ Esin, A. 1998, private communication.
\re
5.\ Garcia M. R., Callanan P. J., McClintock J. E., Zhao P., 1996,
        ApJ, 460, 932
\re
6.\ Gliozzi, M., Bodo, G., Ghisellini, G., Scaltriti, F., Trussoni, E.
1998, A\&A, 337, L39. 
\re
7.\ Hjellming R. M., Rupen M. P., 1995, Nat, 375, 464
\re
8.\ Hynes, R.I,  and Haswell, C.A., 1999, MNRAS, in press, {\tt astro-ph/980941}.
\re
9.\ Hynes, R.I, Haswell, C.A., Shrader, C.R., Chen, W., Horne, K., Harlaftis, E.T.,
O'Brien, K., Hellier, C., and Fender, R.P. 1998a, MNRAS, 300, 64. 
\re
10.\ Hynes, R.I, O'Brien, K.O., Horne, K., Chen, W., and Haswell, C.A. 1998b,
MNRAS, 299, L37.
\re
11.\ Orosz J. A., Bailyn C. D., 1997, ApJ, 477, 876 (OB97)
\re
12.\ Scaltriti F., Bodo G., Ghisellini G., Gliozzi M., and Trussoni E. 1997,
A\&A, 325, 29.
\re
13.\ Shlosman I., Vitello P., 1993, Ap.J, 409, 372                                         

\end{document}